\documentclass{IOS-Book-Article}     
\usepackage{graphicx}
\usepackage{listings}
\usepackage{xcolor}

\usepackage[inline]{enumitem}
\setlist{nolistsep}
\newlist{inlinelist}{enumerate*}{1}
\setlist[inlinelist,1]{label=\textit{\roman*)}}
\newlist{inlineabc}{enumerate*}{1}
\setlist[inlineabc,1]{label=\textit{\alph*)}}

\begin{document}
\pagestyle{plain}
\pagenumbering{arabic}
\begin{frontmatter}
\title{How to Manage My Data? With Machine--Interpretable GDPR Rights!}
\textcolor{red}{\small{Presented at 37th International Conference on Legal Knowledge and Information Systems (JURIX) 2024}}
\author[A]{\fnms{Beatriz} \snm{Esteves}
\thanks{Corresponding authors' emails: beatriz.esteves@ugent.be, me@harshp.com}},
\author[B]{\fnms{Harshvardhan J.} \snm{Pandit}},
\author[C]{\fnms{Georg P.} \snm{Krog}},
\author[B,D]{\fnms{Paul} \snm{Ryan}}
\runningauthor{B. Esteves et al.}
\address[A]{IDLab, Ghent University -- imec, Ghent, Belgium}
\address[B]{ADAPT Centre, Dublin City University, Dublin, Ireland}
\address[C]{Signatu AS, Oslo, Norway}
\address[D]{Uniphar PLC, Dublin, Ireland}

\begin{abstract}
The EU GDPR is a landmark regulation that introduced several rights for individuals to obtain information and control how their personal data is being processed, as well as receive a copy of it. However, there are gaps in the effective use of rights due to each organisation developing custom methods for rights declaration and management.
Simultaneously, there is a technological gap as there is no single consistent standards-based mechanism that can automate the handling of rights for both organisations and individuals.
In this article, we present a specification for exercising and managing rights in a machine-interpretable format based on semantic web standards.
Our approach uses the comprehensive Data Privacy Vocabulary to create a streamlined workflow for individuals to understand what rights exist, how and where to exercise them, and for organisations to effectively manage them.
This work pushes the state of the art in GDPR rights management and is crucial for data reuse and rights management under technologically intensive developments, such as Data Spaces.
\end{abstract}

\begin{keyword}
GDPR, rights management, rights exercise, semantic technologies
\end{keyword}
\end{frontmatter}

\section{Introduction}
\label{sec:intro}

The General Data Protection Regulation (GDPR)~\cite{gdpr} grants data subjects a set of rights designed to protect their personal data and to ensure that they have greater control over how their data is collected, processed, and used by organisations.
These rights include the right to access personal data, the right to rectification of inaccurate data, the right to erasure,
and the right to data portability, which allows individuals to ask for their data to be transferred between data controllers.
Additionally, the GDPR provides rights to restrict processing, object to data use, and avoid automated decision-making.
These rights, combined with the transparency and accountability measures imposed on organisations, aim to strike a balance between the interests of data subjects and the legitimate needs of businesses and institutions in the digital age.
However, despite the comprehensive rights granted to individuals under regulations such as the GDPR, a `technological gap', i.e., a significant lack of efficient tools, for exercising these rights remains~\cite{bernes_enhancing_2022}, with many organisations struggling to provide accessible, user-friendly mechanisms for individuals to manage their personal data, often relying on cumbersome manual processes, which can lead to delays, or even non-compliance.

In this context, we present a specification to express information regarding rights exercise and management in a machine-interpretable format, using semantic Web standards.
In doing so, we begin to address the automation, interoperability and standardisation challenges identified above, towards enabling the development of tools for assisting individuals and organisations in right exercising activities. 
Towards achieving this goal, we reuse and extend the Data Privacy Vocabulary~(DPV)~\cite{panditCreatingVocabularyData2019,pandit2024dpv}, which allows the expression of information related to legislative requirements such as the GDPR, to express information about:
(i) linking personal data processing activities to applicable rights,
(ii) providing notices related to said rights,
(iii) documenting the exercise of rights, and
(iv) GDPR rights requests.
The resulting specification is being developed in the context of the W3C Data Privacy Vocabularies and Controls Community Group (DPVCG).

The remaining of the article is structured as follows: Section~\ref{sec:sota} provides background information on GDPR's data subject rights and existing Web standards for rights management, Section~\ref{sec:rights} presents the developed specification, 
and Section~\ref{sec:conclusions} concludes and presents future directions of work.

\section{Background and State of the Art}
\label{sec:sota}

GDPR's data subject rights are provided on Chapter III, with Articles 12 and 23 listing requirements for and exceptions to the exercising of rights, respectively, and with Articles 13 to 22 concretely defining them.
In essence, these rights involve the flow of information between a data subject and a data controller.
After confirming receipt of the data subject's request, the controller must verify the identity of the data subject before proceeding with it.
If the controller is unable to identify the data subject, additional information must be provided by the data subject to enable identification.
If the controller has a valid justification for not fulfilling the right, the data subject must be informed of such justification.
In cases where the request is complex or there are a large number of requests, the controller is granted a 2-month extension to fulfill the request.
If the controller fails to meet its obligations at any point, a GDPR breach occurs.
The European Data Protection Board also endorsed and issued guidelines to assist on data subject rights~\cite{european_data_protection_board_guidelines_2023,article_29_data_protection_working_party_guidelines_2017,sis_ii_supervision_coordination_group_schengen_2023}.

Even though the existence of these rights has been brought forward has a huge step for individuals empowerment over their personal data, no standards have been defined for the management of rights.
An exception to this statement might be the `Data Rights Protocol'\footnote{\url{https://datarightsprotocol.org}}, which defines a protocol for rights request/response data flows, without however covering all GDPR rights and without using standard vocabularies.
Since semantic technologies promote interoperability, allowing information and its interpretability to be consistently defined across different platforms, 
we will adapt the usage of semantic Web standards for the definition of a specification to express rights exercising and management.
As such, DPV\footnote{\url{https://w3id.org/dpv}; prefixed as  \texttt{dpv}}\cite{panditCreatingVocabularyData2019,pandit2024dpv} is the main driver of the specification presented in this work, as it is the most comprehensive data protection-related vocabulary, providing a set of taxonomies for the expression of metadata related to the usage of personal data aligned with legal requirements, e.g., the GDPR.
Additionally,
ODRL~\cite{iannella_odrl_2018}, 
DCAT~\cite{albertoni_dcat_2024}, 
DCMI~\cite{dcmi_2020}, and 
PROV-O~\cite{lebo_prov_2013}, 
which are Web standards developed and promoted by the W3C, are also used to express information about policies, catalogs of resources, metadata and activity provenance information, respectively.

\section{Rights Exercise and Management}
\label{sec:rights}

Based on the GDPR provisions identified above, the following requirements are addressed in our specification:
\begin{inlineabc}
    \item[(\textit{\ref{sec:applicable-rights}})] information about the existence of rights and justifications for the exercising of said rights, 
    \item[(\textit{\ref{sec:notices}})] notices related to the fulfilment or non-fulfilment of rights,
    \item[(\textit{\ref{sec:records}})] records of rights-related activities, and
    \item[(\textit{\ref{sec:policies}})] rights requests as machine-executable policies.
\end{inlineabc}
The work is publicly available\footnote{Accessible at  \url{https://w3id.org/people/besteves/rights}} and includes modelled examples for each requirement identified above.

\subsection{Applicable Data Subject Rights and Justifications}
\label{sec:applicable-rights}

Beyond jurisdiction-dependence, applicable data subject rights also depend on the legal ground used to process personal data.
In the case of the GDPR, DPV's GDPR extension\footnote{\url{https://w3id.org/dpv/legal/eu/gdpr} prefixed as \texttt{eu-gdpr}} contains the terms to represent the rights available under GDPR, as well as a mapping of which rights are applicable based on the used~legal basis.
Apart from detailing information about what personal data is being processed, how, where, by whom, and for what purpose, a \texttt{dpv:Process} can also be used to indicate applicable rights.
Data controllers can use this approach to express which rights apply, including those beyond GDPR, such as the EU’s fundamental rights and rights outlined in other EU regulations or jurisdictions, i.e., the \texttt{dpv:hasScope} property can be used to indicate applicable rights by jurisdiction.

Moreover, as a result of the collection of requirements for GDPR rights exercising, a \texttt{Justification} taxonomy\footnote{\url{https://w3id.org/dpv/justifications} prefixed as \texttt{justifications}}, to provide reasons or explanations related to the~fulfilment, non-fulfilment, delay and exercise of rights, was developed.
Examples of justifications include individual's identity not being verifiable or a request being considered overly excessive.
Justifications can also explain why certain processes are necessary or being delayed, such as the ground to exercise the right to object or requiring additional information to move forward with a process.
To facilitate the expression of these justifications, this extension introduces concepts that extend DPV's \texttt{Justification} concept.
Moreover, since such concepts can be reused for other data protection-related activities, e.g., data breach reports, the concepts were defined in a generic way so that they can be reused in distinct contexts.

\subsection{Notices related to the fulfilment or non-fulfilment of rights}
\label{sec:notices}

Right exercise notices should inform data subjects about where and how to exercise a right, what information is needed, or provide updates on a submitted right request.
DPV’s \texttt{isExercisedAt} property can be used to link rights to specific exercise points, e.g., through right exercise notices, and to connect the required information with a process using DPV’s \texttt{hasProcess} property.
DPV's \texttt{hasRecipient}, \texttt{hasStatus}, and \texttt{hasJustification} properties can be employed to update the data subject on the status of their right exercising activity, including any justification for why the right is being exercised.
In addition, DCMI's concepts can be reused to indicate temporal and other metadata information related to notices\footnote{DCMI's reused terms are accessible at \url{https://w3id.org/people/besteves/rights\#dcmi}.}.
The notices should also provide information about the data controller, as well as the entity responsible for implementing the notice.


Notices specifically informing about the fulfilment, progress toward fulfilment, or non-fulfilment of a right can also be modelled with DPV.
Actions~required by entities to fulfil a rights-related request, which cannot be fully detailed using DPV-- such as issuing payment terms --can be attached to these notices, e.g.,~using ODRL policies.
For GDPR-specific notices, the GDPR extension offers concepts for both direct and indirect data collection notices, as required by Arts. 13 and 14, and it also includes Subject Access Request notices to fulfil the GDPR Right of Access, and recipient notices to meet the notification requirements of Art.~19.



\subsection{Records of rights-related activities}
\label{sec:records}

Recording provenance information when a specific instance of a right has been or is being exercised is valuable for data controllers, as it tracks information about the status and fulfilment of requests, which can be useful, e.g., for auditors.
To represent specific records of rights being exercised, DPV's \texttt{RightExerciseRecord} can be used to link a particular request, or multiple requests from the same data subject, with the corresponding activities, i.e., \texttt{dpv:RightExerciseActivity}, carried out by entities to fulfil those requests.
These records can also leverage the DCAT standard for representing data catalogues, i.e., \texttt{dcat:Catalog}, with right exercise activities represented as \texttt{dcat:Resource}, which can be grouped and organised using \texttt{dcat:DatasetSeries}.
This approach offers the advantage of utilising DCAT's ordering properties, such as \texttt{dcat:first}, \texttt{dcat:last}, and \texttt{dcat:prev}, to easily track and retrieve the most recent activity related to a specific rights request.

In this context, a \texttt{dpv:RightExerciseActivity} represents a specific instance of an activity carried out in the process of exercising a right.
These activity instances should include metadata, e.g., timestamps and involved entities, to track the provenance of the right exercising process.
To track the status of rights exercising activities, DPV provides \texttt{RequestStatus} concepts, such as \texttt{RequestAccepted} for accepted requests moving toward fulfilment or \texttt{RequestRequiresAction} for requests requiring further action from another party.
Once a request is initiated, it should be acknowledged by the responsible entity and either accepted for fulfilment or rejected.
If rejected, the entity may require additional actions from the requester (e.g., providing more information to proceed), which can delay the acceptance or rejection.
After the required action is taken, the request may either be accepted for fulfilment, rejected again, or further action may be requested.
Moreover, DPV concepts can be combined with PROV-O to track the provenance of a right exercising activity instance.
This allows for the representation of provenance details, such as the entities involved in the activity, i.e., \texttt{prov:wasAssociatedWith}, or the data/notice generated by the activity, i.e., \texttt{prov:generated}.

\subsection{GDPR-related rights requests as machine-executable policies}
\label{sec:policies}

DPV and ODRL can also be used for data subjects to send GDPR-related right requests in a machine-interpretable format to data controllers.
Such requests can then be integrated in their rights management processes for automated execution of responses to data subjects' requests.
For example, systems that use an ODRL evaluator based on the formal semantics of ODRL~\cite{fornara_odrl_2024} can assess policies to determine which ones are active, have been violated, or have been fulfilled, to respond consistently and interoperably to rights-related requests.
As such, GDPR's data subject rights described in Articles 15 to 22 can be instantiated as ODRL policies containing permissions, prohibitions and obligations to data controllers to act upon or fulfil.
Generic policies for each right are available in the specification.

\section{Conclusions and Future Work}
\label{sec:conclusions}

This article provides a machine-interpretable model to represent information regarding data subject rights, 
notices to communicate about them with other entities, records for auditing and policies for automated execution of responses to rights requests based on the GDPR.
We believe our proposal provides a crucially missing technological piece that is required to make effective use of GDPR rights across different services that utilise the Web, in an interoperable manner.
DPV, a central component of our rights management specification, is a comprehensive and growing resource that is being actively developed, has been used by industry and academic projects, and involves inputs from a diverse range of stakeholders.
More importantly, the development of this specification involves engagement with legal experts who validate its use as per legal requirements\footnote{
The author Paul Ryan is a DPO for several companies and the author Georg P. Krog is the chief legal counsel for Signatu -- which uses DPV to provide GDPR compliance solutions.}.


Finally, we highlight the importance of this work in developing ecosystems based on the Data Governance Act (DGA), Data Act, European Digital Identity (EUDI) and Data Spaces regulations, where data reuse, rights and consent management will be based on technical infrastructures.
The work presented in this article provides the technical basis for creating a GDPR-compliant by design rights management infrastructure that Data Spaces or EUDI wallets can utilise for individuals and organisations to exercise and manage rights in a standardised and interoperable manner.
To enable its technical adoption, we will explore the integration of our work with protocols such as the Data Rights Protocol or the Advanced Data Protection Control\footnote{\url{https://www.dataprotectioncontrol.org/spec/}} which have rights management in their scope but lack the technical implementation -- which this work can provide. 

\vspace{2mm}\noindent\textbf{Acknowledgments}
\small{This project has received funding from the EU’s Horizon 2020 research and innovation programme under the Marie Skłodowska-Curie grant agreement No 813497 (PROTECT ITN). Beatriz Esteves is funded by SolidLab Vlaanderen (Flemish Government, EWI and RRF project VV023/10).
The ADAPT Research Centre for AI-Driven Digital Content Technology is funded by Science Foundation Ireland (SFI) under Grant\#13/RC/2106\_P2.}

\bibliography{paper}
\bibliographystyle{vancouver}

\end{document}